\documentclass[acmsmall,screen,nonacm]{acmart}

\usepackage{tikz}
\usetikzlibrary{tikzmark}
\usepackage{hyperref}
\usetikzlibrary{arrows}
\usepackage[utf8]{inputenc}
\usepackage{amsmath}
\usepackage{amsthm}
\usepackage{mathpartir} 
\usepackage{mathtools}
\usepackage{centernot} 
\usepackage{graphicx}
\usepackage{subcaption}
\usepackage{xcolor} 
\usepackage{soul}
\usepackage{float}
\usepackage[font={small,it}]{caption}
\usepackage{import}
\usepackage{calc}  
\usepackage{enumitem}  
\usepackage{environ}
\usepackage{declarations}
\usepackage{stmaryrd}
\usepackage[T1]{fontenc} 
\usepackage{xspace}
\usepackage{listings}
\usepackage{adjustbox}
\usepackage{multirow}
\usepackage{booktabs}
\usepackage{makecell} 

\usepackage{minted}
\usepackage{array}
\usepackage{wrapfig}
\usepackage[most]{tcolorbox}
\usepackage{pifont}
\usepackage{fancyvrb}
\usepackage{threeparttable}

\tcbset{
  mytheorem/.style={
    enhanced,
    colback=red!4!white,
    colframe=red!50!white,
    boxrule=0.6pt,
    arc=2mm,
    outer arc=2mm,
    fonttitle=\bfseries,
    coltitle=black,
  }
}
\tcbset{
  mylemmaobox/.style={
    enhanced,
    colback=blue!3!white,   
    colframe=blue!30!white, 
    boxrule=0.6pt,
    arc=2mm,
    outer arc=2mm,
    fonttitle=\bfseries,
    coltitle=black,
  }
}

\definecolor{lightgray}{rgb}{0.95, 0.95, 0.95}

\usepackage{tikz}
\usetikzlibrary{arrows} 
\usepackage{pstricks-add, pst-pdf}%
\usepackage{pgfplots}
\pgfplotsset{compat=1.18}
\usepackage{colortbl}
\usepackage{tabulary}
\usepackage{etoolbox}
\definecolor{codegreen}{rgb}{0,0.6,0}
\definecolor{codegray}{rgb}{0.5,0.5,0.5}
\definecolor{codepurple}{rgb}{0.58,0,0.82}
\definecolor{backcolour}{rgb}{1, 1, 0.96}
\usepackage{color}
\lstdefinestyle{mystyle}{
  language=Rust,
  backgroundcolor=\color{backcolour},
  commentstyle=\color{codegreen},
  numberstyle=\tiny\color{codegray},
  keywordstyle=\color{magenta},
  stringstyle=\color{codepurple},
  basicstyle=\ttfamily\footnotesize,
  breakatwhitespace=false,
  breaklines=true,
  captionpos=b,
  keepspaces=true,
  numbers=left,
  numbersep=5pt,
  showspaces=false,
  showstringspaces=false,
  showtabs=false,
  tabsize=2
}

\definecolor{comment}{rgb}{0.5, 0.5, 0.5}
\definecolor{keyword}{rgb}{0.7, 0.1, 0.1}
\definecolor{string}{rgb}{0.2, 0.5, 0.2}

\lstdefinelanguage{Rust}{
  morekeywords={pub, struct, impl, trait, use, fn, let, mut},
  sensitive=true,
  morecomment=[l]{//},
  morecomment=[s]{/*}{*/},
  morestring=[b]",
}

\lstdefinelanguage{Rust}{
  morekeywords={pub, struct, enum, fn, impl, let, mut, ref, match, if, else, for, in, while, loop, return, break, continue, Self, super, crate, mod, use, as, where, trait},
  sensitive=true,
  morecomment=[l]{//},
  morecomment=[s]{/*}{*/},
  morestring=[b]{"},
}
\lstdefinestyle{ruststyle}{
  language=Rust,
  commentstyle=\color{codegreen},
  basicstyle=\ttfamily\footnotesize,
  numbers=left,
  numberstyle=\tiny\color{gray},
  stepnumber=1,
  numbersep=8pt,
  showstringspaces=false,
  tabsize=4,
  breaklines=true,
}

\lstdefinestyle{codecase}{ %
    language=Haskell,
    backgroundcolor=\color{pink!20},
    basicstyle=\small\ttfamily,
    keywordstyle=\small\bfseries,
    numberstyle=\small\ttfamily,
    escapeinside={\%*}{*)},
    keywordstyle=\color{blue}
}
\lstdefinestyle{codecaseour}{ %
    backgroundcolor=\color{pink!20},
    basicstyle=\small\ttfamily,
    keywordstyle=\small\bfseries,
    numberstyle=\small\ttfamily,
    keywordstyle=\color{blue}
}
\AtBeginDocument{%
  }

\setcopyright{acmlicensed}
\copyrightyear{2018}
\acmYear{2018}
\acmDOI{XXXXXXX.XXXXXXX}
\acmConference[Conference acronym 'XX]{Make sure to enter the correct
  conference title from your rights confirmation email}{June 03--05,
  2018}{Woodstock, NY}
\acmISBN{978-1-4503-XXXX-X/2018/06}

\begin{document}

\title{\toolname: Denning-Style Information Flow Control for Rust}

\author{Jeffrey C. Ching}
\affiliation{%
  \institution{Duke University}
  \country{USA}
}
\author{Quan Zhou}
\affiliation{%
  \institution{Google}
  \country{USA}
}
\author{Danfeng Zhang}
\affiliation{%
  \institution{Duke University}
  \country{USA}
}
\authorsaddresses{Authors’ Contact Information: Jeffrey C. Ching, Duke University, Durham, NC, USA, jcc150@duke.edu; Quan Zhou, Google, USA, quanzhour@google.com; Danfeng Zhang, Duke University, Durham, NC, USA, dz132@duke.edu}
\renewcommand{\shortauthors}{Jeffrey C. Ching, Quan Zhou, and Danfeng Zhang}

\begin{abstract}
Existing language-based information-flow control (IFC) tools face a fundamental tension: Denning-style systems that track explicit and implicit flows at the variable level typically require compiler modifications, while more coarse-grained approaches, including recent work Cocoon, avoid compiler changes but impose more restrictive programming models.

We present \toolname, a Denning-style static IFC library for Rust that requires no compiler modifications. \toolname addresses three key challenges in building a practical IFC library for Rust. First, it enables fine-grained explicit-flow checking with minimal annotation overhead by leveraging Rust’s type inference. Second, it introduces \lstinline|pc_block!|, a lightweight construct for enforcing implicit flows via a compile-time program counter label, without requiring compiler support. Third, it provides \lstinline|fcall!| and \lstinline|mcall!| macros to support seamless and safe interoperability with standard and third-party libraries.
Our evaluation shows that \toolname incurs negligible compile-time overhead and requires only modest annotations. Moreover, compared to Cocoon, \toolname offers a more permissive programming model, reducing the need for frequent escape hatches that bypass security checks.

\end{abstract}



\maketitle
\definecolor{codebg}{HTML}{F7F7F7}
\definecolor{kw}{HTML}{7B2C8B}
\definecolor{cmt}{HTML}{6A9955}
\definecolor{str}{HTML}{A31515}
\definecolor{mac}{HTML}{0070C1}
\definecolor{hel}{HTML}{e51284}

\lstdefinelanguage{RustIFC}{
  language=Rust,
  keepspaces=true,
  alsoletter={!},
  morekeywords={fn,let,mut,pub,struct,impl,unsafe,trait,for,if,else,
                while,return,use,where,type,Self,self,async,await,
                move,true,false,Some,None,Ok,Err},
  morekeywords=[2]{Labeled,Public,A,B,AB,Label,FlowsTo,Join,
                   PcContext,Vetted,InvisibleSideEffectFree},
  morekeywords=[3]{fcall!,mcall!,pc_block!,relabel!,declassify},
  morekeywords=[4]{side_effect_free_attr, check_isef, unchecked_operation},
  keywordstyle=\color{kw}\bfseries,
  keywordstyle=[2]\color{blue!70!black},
  keywordstyle=[3]\color{mac}\bfseries,
  keywordstyle=[4]\color{hel}\bfseries,
  commentstyle=\color{cmt}\itshape,
  stringstyle=\color{str},
  basicstyle=\ttfamily\footnotesize,
  backgroundcolor=\color{codebg},
  frame=single,
  rulecolor=\color{black!20},
  xleftmargin=2em,
  numbers=left,
  numberstyle=\tiny\color{gray},
  numbersep=6pt,
  tabsize=4,
  showstringspaces=false,
  breaklines=true,
  captionpos=b,
  mathescape=false,
  literate={->}{{$\rightarrow$}}2
           {=>}{{$\Rightarrow$}}2
           {<=}{{$\sqsubseteq$}}1,
}

\lstset{language=RustIFC}
\lstset{escapeinside={(*@}{@*)}}

\section{Introduction}

Information flow control (IFC) is a fundamental mechanism for ensuring that sensitive data is accessed and transmitted only in accordance with established security policies. By regulating how information propagates through a system, it helps prevent unauthorized disclosure and supports the preservation of confidentiality, integrity, and privacy. Denning's seminal work~\cite{denning1977}, which we term as \emph{Denning-style IFC}, introduced a precise, fine-grained framework that tracks information flow through both explicit assignments and implicit control dependencies using a lattice model of security classes~\cite{denning1976}. It is characterized by:
\begin{itemize}
    \item A program counter ($\pc$) label, which tracks the security level of the current control-flow context, in order to control implicit flows introduced by branches and loops.

    \item IFC checks at \emph{each instruction} to control both explicit and implicit flows.  
\end{itemize}
Compared with coarse-grained IFC systems such as those based on the Bell–LaPadula model~\cite{bell1973}, which enforce a simple lattice-based policy known as ``no read up, no write down,'' at the granularity of entire subjects and objects (e.g., processes and files), Denning-style IFC offers increased precision (fewer unnecessary rejections of safe programs) via reasoning about information propagation \emph{within} complex software systems, as well as provable end-to-end guarantees about confidentiality and integrity~\cite{volpano1996, pottier2002, Sabeleld2003}. This strong theoretical basis has enabled practical language implementations such as Jif~\cite{myers1999, myers1997, jif-cornell} and its variants~\cite{liu2009fabric, clarkson2008civitas}. 

However, due to the inherent complexity of Denning-style IFC, existing implementations typically require modifications to both the programming language and its compiler, including customized type-checking and inference rules. These requirements significantly hinder the practical adoption of these well-founded techniques. In particular, it is both costly and time-consuming to rewrite large existing software systems in specialized languages and to maintain customized compilers to keep pace with their underlying base language features.

These limitations have motivated a promising line of recent work~\cite{russo2008library, russo2015functional, gregersen2019dependently, lamba2024, beardsley2025carapace} that investigates how IFC can be enforced through lightweight libraries layered on top of mainstream programming languages. For example, Russo et al.~\cite{russo2008library} propose a \emph{monadic} library for Haskell in which confidential values are encapsulated within an abstract security type, $\cod{Sec}$. By construction, the only way to manipulate secret values is through the monadic operations provided by the library. 
%
Because such effective systems are generally not supported by mainstream imperative languages such as Java, C, and Rust, a recent work Cocoon~\cite{lamba2024} introduces the first static IFC mechanism for the Rust language. Notably, Cocoon ensures that all reads and writes involving secret values occur within lexically-scoped regions, called \emph{secret blocks}, which are introduced via the $\cod{secret\_block}$ macro. For example, the Cocoon code snippet below performs a computation over $\cod{x}$ and $\cod{y}$ and stores the result in $\cod{z}$, where $\cod{x}$ and $\cod{z}$ are secret while $\cod{y}$ is public:

\begin{lstlisting}[label={lst:cocoon-example}]
#[side_effect_free_attr]
fn foo(x: i32, y: i32) -> i32 { x + y }
let x = secret_block!(lat::A { wrap_secret(1) });
let y = 2;
let z: Secret<i32, lat::A> = secret_block!( lat::A {
    let u_x = unwrap_secret(x);
    wrap_secret(foo(u_x, y)) 
});
\end{lstlisting}

In Cocoon, non-public values are wrapped in the abstract type $\cod{Secret}$, e.g., $\cod{Secret\langle i32, lat::A \rangle}$ denotes an integer classified at level $\cod{A}$. Accessing or modifying such values requires execution within a secret block (e.g., lines 4 and 7–8), the only lexically scoped region in which non-public values may be revealed using the macro $\cod{unwrap\_secret}$ (e.g., line 7). Cocoon enforces several restrictions to preserve security guarantees. For a secret block with level $\cod{L}$, the block may (1) unwrap values whose security level is \emph{no higher than} $\cod{L}$ (e.g., line 7); (2) modify non-local variables whose security level is \emph{no lower than} $\cod{L}$; (3) return values classified \emph{exactly at} level $\cod{L}$ (e.g., line 8); and (4) invoke only side-effect-free functions, i.e., functions that do not write to variables visible outside their local scope (e.g., function $\cod{foo}$, annotated at line 1). Since all of those restrictions are applied to the entire secret block, we term such an enforcement \emph{block-level IFC}, to contrast with the Denning-style IFC that tracks $\pc$ labels and checks information flows for each individual expression and command.




In this paper, we present \toolname, the first Denning-style
IFC system for a mainstream imperative language without changing the compiler. Similar to Cocoon and its successor Carapace~\cite{beardsley2025carapace}, \toolname is implemented as a Rust library for unmodified Rust and operates with the standard Rust compiler. However, \toolname advances beyond Cocoon’s block-level protection by \emph{weaving fine-grained IFC directly at the level of individual instructions} via the following novel features:
\begin{itemize}
\item
Rather than unwrapping (i.e., downgrading) non-public values within secret blocks, \toolname employs a \lstinline|relabel!| macro to explicitly \emph{upgrade} sensitivity when necessary. Consequently, secret values are not required to remain confined within designated secret blocks\footnote{By eliminating the secret blocks used in Cocoon, whose implementation relies on auto traits and negative traits not available in stable Rust, \toolname remains compatible with stable Rust, which is an additional benefit of its novel design.}.

\item In contrast to conservatively prohibiting all side effects within a secret block, \toolname introduces a \lstinline|pc_block!| construct to track the $\pc$ label. Under this mechanism, only side effects whose sensitivity is lower than the current $\pc$ are disallowed. As a result, \toolname eliminates the need for the no-side-effect requirement check for majority of functions.

\item To make it compatible with the Rust standard library and third-party crates, \toolname leverages Rust’s type system and introduces novel macros, such as \lstinline|fcall!| and \lstinline|mcall!|. These macros enable labeled values to be passed through unlabeled function calls without requiring declassification, while soundly tracking label propagation across library boundaries.

\item \toolname takes advantage of the Rust trait system and type inference system to reduce the annotation burden on programmers. As a result, \toolname avoids frequent use of annotations such as $\cod{unwrap\_secret}$ and $\cod{wrap\_secret}$ in Cocoon programs.

\end{itemize}

Together, the metaprogramming layer introduced by \toolname effectively bridges the gap between Denning-type IFC and Rust's existing type system, allowing IFC on top of a stable Rust compiler. For example, the following \toolname code is equivalent to Cocoon code above:
\begin{lstlisting}[label={lst:fcall-example}]
fn foo(x: i32, y: i32) -> i32 { x + y }
let x = Labeled::<i32, A>::new(1);
let y = 2;
let z = fcall!(foo(x, relabel!(y, A)));
\end{lstlisting}

We note that \toolname provides the following benefits due to its novel features described above:
\begin{itemize}
    \item \toolname eliminates the need for secret blocks, side-effect checks, and explicit wrapping and unwrapping operations present in the Cocoon equivalent. As a result, \toolname code more closely resembles the original Rust code written without IFC mechanisms.

    \item When invoking the function $\cod{foo}$ with a non-public parameter $\cod{x}$, the macro \lstinline|fcall!| automatically lifts the parameter and returns types of $\cod{foo}$ to the secret type \lstinline|Labeled::<i32, A>|. Additionally, the macro \lstinline|relabel!| is used to upgrade $\cod{y}$ to the same security level.

    \item Leveraging Rust’s type inference system, \toolname requires only minimal type annotations.
\end{itemize}

We implement \toolname as a Rust library built on Rust 1.69 and have open-sourced it at~\cite{implementation}. To evaluate its practicality, we re-implemented all four applications from the Cocoon code repository~\cite{cocoon-git}, along with an additional application, JPMail~\cite{hicks2006}, originally developed in Jif.
Our evaluation yields several important findings. First, \toolname enforces security guarantees comparable to prior approaches while requiring only minimal to moderate code modifications across all applications. Second, although \toolname adopts a fine-grained, Denning-style IFC model, the annotation burden remains low in practice due to its design, which leverages Rust’s type inference mechanisms. Third, \toolname provides a more permissive programming model than Cocoon: it eliminates 12 unnecessary declassification and unchecked operations present in Cocoon's implementations, thereby reducing potential risks for unintended information leakage. Finally, we observe that the compile-time overhead with \toolname is low and scales well to large codebases.

In summary, this paper makes the following contributions:
\begin{itemize}
    \item We develop the first Denning-style information flow tool in Rust without requiring compiler modifications.

    \item We design \toolname to address key challenges in Denning-style IFC for Rust by enabling fine grained information flow tracking with low annotation burden, supporting implicit flow tracking via \lstinline|pc_block!| without compiler changes, and ensuring seamless and safe interoperability with external libraries through \lstinline|fcall!| and \lstinline|mcall!|.
    
    \item Compared to Cocoon, \toolname requires less code modification and is more faithful to the original program. Moreover, the permissive programming model of \toolname also removes unnecessary declassification in Cocoon code.
    
    \item We present a performance evaluation demonstrating that \toolname's low compile-time overhead and minimal annotation burden.
\end{itemize}

\section{Background and Overview}
\label{sec:background}

\subsection{Secrecy Labels and Lattice}
\label{sec:lattice}

As is standard in IFC systems, we associate all data with security labels dawn from a predefined security lattice $\mathcal{L}$. Two labels $\ell_1$ and $\ell_2$ from the lattice are ordered, written $\ell_1 \sqsubseteq \ell_2$, if information at $\ell_2$ is at least as restrictive as information at $\ell_1$ (in other words, data labeled with $\ell_1$ can safely flow to $\ell_2$). In a confidentiality lattice, $\ell_1 \sqsubseteq \ell_2$ means that $\ell_2$ is more sensitive than $\ell_1$, while in an integrity lattice, $\ell_1 \sqsubseteq \ell_2$ means that $\ell_1$ is more trusted than $\ell_2$. Without loss of generality, we assume a ``diamond'' lattice with four labels:
$\bot$ (public), $\ell_A$, $\ell_B$, and $\ell_{AB}$, ordered as $\bot \sqsubseteq \ell_A \sqsubseteq \ell_{AB}$ and
$\bot \sqsubseteq \ell_B \sqsubseteq \ell_{AB}$ throughout the paper. The join operation $\ell_1 \sqcup \ell_2$ returns the least upper bound of two labels in the lattice, representing the most permissive label that is at least as restrictive as both $\ell_1$ and $\ell_2$. For example, $\ell_A \sqcup \ell_{AB}= \ell_{AB}$ and $\ell_A \sqcup \ell_{B}= \ell_{AB}$.

\subsection{Denning-Style IFC}
\label{sec:denning}

Early IFC systems, such as those based on the Bell--LaPadula model~\cite{bell1973}, enforced simple lattice-based policies commonly summarized as ``no read up, no write down.'' These policies regulate information flow at the granularity of entire subjects and objects (e.g., processes and files). For instance, a process with label $\ell_A$ cannot read data labeled $\ell_{AB}$, nor can it write to data labeled \textit{public}. Such enforcement mechanisms were primarily deployed in OS-level IFC systems~\cite{efstathopoulos2005labels, zeldovich2008securing}, as well as in dynamic IFC systems that raise security labels at runtime when sensitive data is accessed.
However, under this model, once confidential data enters the execution context, the entire computation becomes effectively ``tainted.'' As a result, all subsequent outputs must be treated at the elevated security level associated with that data. This limitation leads to the well-known \emph{label creep} problem~\cite{stefan2011, vassena2019fine}, in which the security level of a long-running process monotonically increases over time. Eventually, the process becomes so highly tainted that it can no longer produce outputs that are observable at lower security levels, such as public outputs.

The seminal work of Denning~\cite{denning1977} laid the foundation for modern fine-grained, language-based IFC systems~\cite{Sabeleld2003}. A key contribution of Denning’s model is the introduction of a \emph{program counter} label ($\pc$), which captures the sensitivity of the current control-flow context.
The $\pc$ label is raised when program execution enters a branch whose condition
depends on sensitive data. This mechanism enables IFC systems to precisely track
implicit flows through control dependencies while avoiding the label creep
problem: once execution exits the scope of the sensitive branch, the $\pc$ label
can be restored to its previous label, thereby allowing subsequent computations
to proceed without unnecessarily propagating high-security labels. For example, consider the following code:
\begin{lstlisting}[language=Rust]
let dataA: i32 = 10;      // with label A, non-public
let mut public: i32 = 0;
public = dataA;           // insecure explicit flow
if dataA > 0 {            // pc is set to label A
    public = 1;           // insecure implcit flow
} else {
    public = 2;           // insecure implicit flow
}                         // pc is set to public
public = 2;               // secure explicit flow
\end{lstlisting}
In this example, the variable \cod{dataA} carries a non-public security label
$\ell_A$, while \cod{public} represents a public variable. The assignment at line 3 constitutes an \emph{explicit flow}, as information
from a non-public source is directly written into a public variable.
The branch at line 4 illustrates an \emph{implicit flow}. Since the branch
condition depends on the non-public variable \cod{dataA}, by observing if $\cod{public}$ is set to 1 or 2 after the branch, an attacker can reveal if \cod{dataA} is greater than zero or not.

To eliminate both explicit and implicit insecure information flows, a Denning-style IFC system associates each program point with a program counter label ($\pc$) that represents the sensitivity of the current control context. For every assignment statement of the form $x := e$, the system enforces two constraints.
First, to prevent explicit flows, the security label of the expression $e$ must be bounded by the label of the target variable $x$ (e.g., $\ell_A \sqsubseteq \bot$ at line 2, which does not hold). Second, to prevent implicit flows, the program counter label $\pc$ must also be bounded by the label of $x$ (e.g., $\ell_A \sqsubseteq \bot$ at lines 5 and 6, which does not hold).
Note that $\pc$ is reset to public at line 9, after the sensitive branch. Consequently, the assignment at line 10 yields the constraints $\bot \sqsubseteq \bot$ for both explicit and implicit flow checks. These constraints trivially hold, and a Denning-style system therefore correctly concludes that the statement at line 10 is secure.

Denning-style IFC can be formalized as a type system that provides provable end-to-end guarantees for both confidentiality and integrity~\cite{volpano1996, pottier2002, Sabeleld2003}. This strong theoretical foundation has enabled the development of practical language-based implementations, including Jif~\cite{myers1999, myers1997, jif-cornell} and several of its extensions and variants~\cite{liu2009fabric, clarkson2008civitas}.

\subsection{Cocoon and Block-Level IFC}
\label{sec:blocklevel}

Cocoon~\cite{lamba2024} represents an important step toward integrating static IFC into mainstream imperative programming languages. In Cocoon, programmers encapsulate sensitive data within a \lstinline|secret_block!|, within which the system enforces the classical information-flow policy of ``no read up, no write down.'' By design, Cocoon is more permissive than coarse-grained IFC systems in two key respects:
\begin{itemize}
    \item Cocoon enforces IFC within the scope of individual secret blocks, rather than across an entire process or function.

    \item Cocoon mitigates the label creep problem by restricting all uses of non-public values to secret blocks and by either verifying or assuming that functions invoked within these blocks are side-effect free. Consequently, the execution environment remains public outside the boundaries of the secret blocks.
\end{itemize}

However, unlike Denning-style IFC systems, Cocoon does not maintain a $pc$ label, nor does it enforce explicit and implicit flows by checking IFC constraints on the security labels of expressions. Instead, Cocoon unwraps non-public data within each secret block, verifies that the functions invoked inside the block are side-effect free, and taints the resulting outputs with the secrecy label associated with the enclosing secret block. Accordingly, we characterize Cocoon's approach as enforcing \emph{block-level IFC} in this paper.

\subsection{Rust}

Rust is a modern programming language designed to provide strong safety guarantees without sacrificing performance. Its expressive type system, featuring traits, generics, lifetimes, and ownership semantics, enables the compiler to enforce rich correctness properties at compile time while still supporting low-level control over memory and concurrency. In the following, we introduce several Rust features on which \toolname builds on.

\subsubsection*{Traits} 
Rust’s trait system provides a mechanism for defining shared behavior abstractly, allowing types to implement specified interfaces and enabling polymorphism and code reuse through trait bounds and generic programming. For example, the standard library defines the \lstinline|Display| trait for types that can be formatted as user-facing strings:
\begin{lstlisting}[language=Rust]
use std::fmt::Display;
fn print_value<T: Display>(x: T) {
    println!("{}", x); }
\end{lstlisting}
Here, the generic function \lstinline|print_value| can accept any type that implements the \lstinline|Display| trait, such as \lstinline|i32| or \lstinline|&str|.

\toolname leverages Rust's trait system to support user-defined security lattices (Section~\ref{sec:lattice}), enable static checking of flows-to relations over security labels, and compute the least upper bound (i.e., join operation) of these labels (Section~\ref{sec:labels}).

\subsubsection*{Generics and Phantom Types}

Rust's generics allow types and functions to be parameterized over arbitrary types subject to trait bounds. For example, 
\begin{lstlisting}[language=Rust]
use std::fmt::Display;
fn print_pair<T: Display>(x: T, y: T) {
    println!("({}, {})", x, y); }
\end{lstlisting}

Here, the function \lstinline|print_pair| is generic over the type parameter \lstinline|T|, while the trait bound \lstinline|T: Display| requires that any instantiation of \lstinline|T| implement the \lstinline|Display| trait.

\toolname uses Rust generics to wrap values of any data type to an abstract security  labeled values (Section~\ref{sec:labeleddata}) and also utilize generics to lift label-less functions to handle labeled values (Section~\ref{sec:funccall}). 

Moreover, Rust supports phantom types through the \lstinline|PhantomData| marker type, which allows programmers to associate a generic type parameter with a data structure without storing a value of that type. This pattern is commonly used to encode additional compile-time information without affecting the runtime representation.
To enforce Denning-style IFC without incurring runtime penalties, we leverage Rust's phantom type to associate security labels with their corresponding data into the type system (Section~\ref{sec:labeleddata}).


\section{Motivating Examples}
\label{sec:Motivation}

As discussed in Section~\ref{sec:background}, Cocoon enforces block-level IFC, whereas \toolname adopts Denning-style IFC. In theory, fine-grained and coarse-grained IFC have been shown to be equally expressive in both static~\cite{rajani2017type, rajani2020} and dynamic~\cite{vassena2019fine} settings. Nevertheless, we draw on several concrete code examples from Cocoon’s repository~\cite{cocoon-git} to illustrate that Denning-style IFC provides a more appealing programming model for developers.

Specifically, although it is theoretically possible to \emph{rewrite} Cocoon programs to overcome the limitations discussed below, our examples suggest that developing IFC-enabled programs in \toolname is more intuitive and natural for programmers. Moreover, the resulting code remains closer to standard Rust code written without IFC, making it easier to port existing Rust projects to \toolname.

\subsection{Annotation Burden and Unchecked Functions}
\label{sec:uncheckedfunction}
\begin{figure}[t]
\centering

\begin{subfigure}[b]{1.0\textwidth} 
\begin{lstlisting}
impl<L: lat::Label> Player<L> {
  fn new() -> Player<L> {
    (*@\highlightlineCo{let ship\_positions = secret\_block!(L \{}@*)let ship_positions = secret_block!(L {
      let ships = [...];
      let mut ship_positions: Grid<bool> = [[false; GRID_SIZE]; GRID_SIZE];
      for ship in ships {
        let placement = random_placement(&ship_positions, &ship);
        place_ship(&mut ship_positions, &placement); 
      }(*@\highlightlineCo{wrap\_secret(ship\_positions)}@*)wrap_secret(ship_positions) });
      ... 
}}

(*@\highlightlineCo{\#[side\_effect\_free\_attr]}@*)#[side_effect_free_attr]
fn random_placement(grid: &Grid<bool>, ship: &Ship) -> Placement

(*@\highlightlineCo{\#[side\_effect\_free\_attr]}@*)#[side_effect_free_attr]
fn place_ship(grid: &mut Grid<bool>, placement: &Placement) {
    ... 
    while row<placement.start_row+placement.size && col<placement.start_col+placement.size {
        (*@\highlightlineCo{unchecked\_operation(grid[row][col] = true);}@*)unchecked_operation(grid[row][col] = true);
        ...
}}
\end{lstlisting}
\vspace{-5pt}
\caption{Implementation of Battleship guess in Cocoon.}
\vspace{5pt}
\label{fig:check_guess_c}
\end{subfigure}
\begin{subfigure}[b]{1.0\textwidth} 
\begin{lstlisting}[escapeinside={(*@}{@*)}]
impl<L: Label> Player<L> {
  pub fn new() -> Player<L> {
    let ships: [Ship; 5] = [ ... ];
    (*@\highlightline{let mut ship\_positions = Labeled::<Grid<bool>,L>::new([[false; GRID\_SIZE];GRID\_SIZE]);}@*)let mut ship_positions = Labeled::<Grid<bool>, L>::new([[false; GRID_SIZE]; GRID_SIZE]);
    for ship in &ships {
      let placement = random_placement(&ship_positions, ship);
      place_ship(&mut ship_positions, &placement);
    }
    ...    
}}

fn place_ship<L: Label>(grid: &mut Labeled<Grid<bool>, L>, placement: &Placement) {
    ...
    grid[row][col] = Labeled::<bool, L>::new(true);
    ...
}}
\end{lstlisting}
\vspace{-5pt}
\caption{Implementation of Battleship guess in \toolname.}
\label{fig:check_guess_f}
\end{subfigure}
\caption{The implementation of player function of battleship game in \toolname and Cocoon.}
\end{figure}

Figure~\ref{fig:check_guess_c} presents a code snippet from Cocoon’s implementation of the Battleship game, a classic two-player board game in which each player places ships at secret locations on a grid, and the placement of each player’s ships must remain confidential from the opponent. Due to its conceptual simplicity coupled with non-trivial security requirements, Battleship has been widely used as a case study in prior IFC systems~\cite{jif-cornell, kozyri2019jrif}.

To protect the confidentiality of $\cod{ship\_positions}$, the initialization logic must be enclosed within a \lstinline|secret_block!| (lines 3–9). Furthermore, every function invoked within this block, including \lstinline|random_placement| (line 14) and \lstinline|place_ship| (line 17), must be individually annotated with \lstinline|#[side_effect_free_attr]|. This manual annotation requirement increases with the depth of the call chain, since any function invoked by an annotated function must also be free of side effects.

Beyond the annotation burden, a more significant limitation is that \emph{disallowing all side effects can be overly restrictive}. For instance, at line 20, the function $\cod{place\_ship}$ must update external data stored in $\cod{grid}$, which inherently introduces side effects. Cocoon’s current workaround is to wrap such operations in \lstinline|unchecked_operation|. However, this mechanism effectively bypasses IFC and may compromise its end-to-end soundness guarantees in such cases.


To address these limitations, \toolname tracks information flow at the level of individual variables, as shown in Figure~\ref{fig:check_guess_f}. Only the secret variable $\cod{ship\_positions}$ is annotated with \lstinline|Labeled<Grid<bool>, L>|, while no additional annotations are required due to Rust’s type inference system.
Moreover, within the function $\cod{place\_ship}$, \toolname can precisely determine that updates to $\cod{grid}$ are secure, since the corresponding parameter $\cod{ship\_positions}$ carries a secret label. Hence, the \toolname version is also more secure, as no IFC bypassing operations like \lstinline|unchecked_operation| are required.

\subsection{Implicit Flows}
\begin{figure}
    \centering
        \begin{subfigure}[b]{1.0\textwidth}
    \begin{lstlisting}
let alice_cal: HashMap<String, Secret<lat::A,bool>> = // { "Monday" -> Secret(true), ... }
let bob_cal: HashMap<String, Secret<lat::B,bool>> =   // { "Monday" -> Secret(false), ...}
let mut count = secret_block!(lat::AB { wrap_secret(0) });
for (day, available) in alice_cal {
    (*@\highlightlineCo{secret\_block!}@*)secret_block!(lat::AB {
      if (*@\highlightlineCo{unwrap\_secret}@*)unwrap_secret(available) &&
          (*@\highlightlineCo{*unwrap\_secret\_mut\_ref}@*)*unwrap_secret_mut_ref(...Option::unwrap(...HashMap::get(&bob_cal, &day))) {
          (*@\highlightlineCo{*unwrap\_secret\_mut\_ref}@*)*unwrap_secret_mut_ref(&mut count) += 1;
      }});}
\end{lstlisting}
\vspace{-5pt}
\caption{Implementation of Calendar in Cocoon.}
\vspace{5pt}
    \label{fig:calendar_cocoon}
    \end{subfigure}
    \begin{subfigure}[b]{1.0\textwidth}
    \begin{lstlisting}
let alice_cal: HashMap<String, Labeled<lA,bool>> = // { "Monday" -> Secret(true), ... } 
let bob_cal: HashMap<String, Labeled<B,bool>> =   // { "Monday" -> Secret(false), ...} 
let mut count = Labeled::<i32, AB>::new(0);
for (day, available) in alice_cal {
  if let Some(bob_avail) = bob_cal.get(day) {
    (*@\highlightline{pc\_block!}@*)pc_block!{(AB) {
      if *available && *bob_avail {
        count = count + 1;
      }}}};}
\end{lstlisting}
\vspace{-5pt}
\caption{Implementation of Calendar in \toolname.}
    \label{fig:calendar_f}
    \end{subfigure}
    \caption{Implicit flow in the calendar example.}
    \label{fig:calendar_example}
\end{figure}

Figure~\ref{fig:calendar_cocoon} presents a code snippet from Cocoon’s implementation of a calendar program that determines the mutual availability of Alice and Bob, where each person’s calendar is considered confidential. Because Cocoon does not distinguish implicit flows from explicit flows, the entire branch between lines 6 and 9 must be enclosed within \lstinline|secret_block!|, as the branch condition depends on Alice’s availability.
As discussed in Section~\ref{sec:uncheckedfunction}, code within the secret block requires frequent use of the \lstinline|unwrap| annotation to perform computations on non-public values. In addition, all function calls within the block, including $\cod{Option::unwrap}$ and $\cod{HashMap::get}$, must either be verified to side-effects free or be wrapped with \lstinline|unchecked_operation| to bypass IFC enforcement.

In contrast, \toolname introduces a novel macro, \lstinline|pc_block!|, to track $\pc$ labels at compile time, as illustrated in Figure~\ref{fig:calendar_f}. At line 5, the $\pc$ label is annotated as \lstinline|lat::AB| because the control flow depends on both Alice’s availability (line 4) and Bob’s availability (line 5). Within a \lstinline|pc_block!| environment, \toolname statically verifies that all modified variables have labels that are at least as restrictive as \lstinline|lat::AB|, thereby preventing illegal implicit flows. For example, if the label of $\cod{count}$ were changed to \lstinline|lat::A|, the assignment to $\cod{count}$ at line 8 would be rejected by the type system. We provide further details on this mechanism in Section~\ref{sec:implicit}.


\subsection{Library Function Calls}
\label{sec:func_calls_mot}
\begin{figure}[t]
  \begin{subfigure}[t]{\columnwidth}
    \begin{lstlisting}[ label={lst:cocoon-set-device}]
let serde_yaml_str = serde_yaml::from_str(config_string.(*@\highlightlineCo{declassify\_ref()}@*)declassify_ref())?;
let mut config_plain = (*@\highlightlineCo{secret\_block!}@*)secret_block!(lat::Label_A { (*@\highlightlineCo{wrap\_secret}@*)wrap_secret(serde_yaml_str) });
self.device_id = Some(device_id.clone());

(*@\highlightlineCo{secret\_block\_no\_return!}@*)secret_block_no_return!(lat::Label_A {
  let u = (*@\highlightlineCo{unwrap\_secret\_mut\_ref}@*)unwrap_secret_mut_ref(&mut config_plain);
  u.device_id = Some(device_id);
});
let new_config = serde_yaml::to_string(config_plain.(*@\highlightlineCo{declassify\_ref()}@*)declassify_ref())?;
write!(config_file, "{}", new_config.(*@\highlightlineCo{declassify\_ref()}@*)declassify_ref())?;
\end{lstlisting}
\vspace{-5pt}
 \caption{Implementation of Spotify TUI in Cocoon.}
\vspace{5pt}
 \label{fig:set-device-id-c}
  \end{subfigure}
  \begin{subfigure}[t]{\columnwidth}
    \begin{lstlisting}[label={lst:ours-set-device}]
let mut config_plain = fcall!(serde_yaml::from_str::<ClientConfigSerde>(&config_string)?);
self.device_id = Some(device_id.clone());
config_plain = mcall!(config_plain.with_device_id(device_id));
let new_config = fcall!(serde_yaml::to_string(&config_plain)?);
mcall!(config_file.write(&new_config)?);
\end{lstlisting}
\vspace{-5pt}
    \caption{Implementation of Spotify TUI in \toolname.}
    \label{fig:set-device-id-f}
\end{subfigure}
  \caption{Comparison of \lstinline|set_device_id| function. Cocoon requires \lstinline|declassify_ref()| at every serde and I/O call boundary, and a \lstinline|secret_block_no_return!| with \lstinline|unwrap_secret_mut_ref| to mutate a single field. Our library uses \lstinline|fcall!| and \lstinline|mcall!| to thread the label through all calls.}
  \label{fig:set-device-id}
\end{figure}


Figure~\ref{fig:set-device-id-c} presents a code snippet from Cocoon’s implementation of Spotify TUI, a terminal based Spotify client written in Rust. This application processes user credentials, including user IDs and passwords. So any information derived from or related to the password is treated as confidential.

The application relies extensively on third party libraries. However, functions provided by these libraries, such as \lstinline|serde_yaml::from_str|, are security-agnostic, meaning they expect plain Rust types rather than security labeled types. To accommodate this limitation, Cocoon requires programmers to (1) explicitly declassify inputs at each function call (lines 1, 9, and 10) in order to invoke these third party functions, and then (2) rewrap the returned values immediately afterward (line 2).
However, this approach introduces a security risk, as it relies on programmers to manually restore the security labels after each call. If the programmer forgets to rewrap a returned value, the type system cannot detect the resulting vulnerability. For example, if the re-wrapping at line 2 is omitted and the code instead is
\lstinline|let mut serializable_config = serde_yaml_str;|
Cocoon fails to detect the error, and the variable $\cod{serializable\_config}$ may subsequently be leaked to the public!



To overcome this limitation, \toolname provides two macros, \lstinline|fcall!| and \lstinline|mcall!|, which automatically lift security-agnostic functions and methods to their labeled variants. As a result, the outputs of these functions and methods are automatically wrapped with the appropriate security label, avoiding the security flaw present in Cocoon’s approach. For example, in \toolname’s implementation shown in Figure~\ref{fig:set-device-id-f}, \lstinline|config_string| can be processed as a value of type \lstinline|Labeled<String, A>| through \lstinline|fcall!| without any manual declassification, and the returned value is lifted to the same security label. Further details of this mechanism are provided in Section~\ref{sec:funccall}.

\section{\toolname Design}

This section presents \toolname, a Denning-style static IFC system built on top of Rust. We begin by describing the program model underlying \toolname in Section~\ref{sec:program_model}. Section~\ref{sec:explicit} explains how \toolname handles explicit information flows, while Section~\ref{sec:funccall} discusses its treatment of function calls. Finally, Section~\ref{sec:implicit} describes how \toolname handles implicit flows. 
The new language constructs are summarized in Figure~\ref{fig:programming-model}.

\begin{figure}
\centering
\small
\renewcommand{\arraystretch}{1.3}

\begin{tabular}{@{}p{3.2cm}p{3.8cm}p{6.5cm}@{}}
\toprule
\textbf{Construct} & \textbf{Syntax} & \textbf{Definition} \\
\midrule
\multicolumn{3}{@{}l}{\textit{Creating \& transforming labeled values}} \\
\addlinespace[2pt]
Wrap
  & \lstinline|Labeled::<T,L>::new(v)|
  & Create \lstinline|Labeled<T,L>| \\
Relabel
  & \lstinline|relabel!(x, Target)|
  & Upgrade label; compile error if $L_x \not\sqsubseteq \mathit{Target}$ \\
Declassify
  & \lstinline|declassify(x)|
  & Downgrade to \lstinline|Public| (escape hatch) \\
\midrule
\multicolumn{3}{@{}l}{\textit{Function Calls \& Method calls}} \\
\addlinespace[2pt]
Function call
  & \lstinline|fcall!(f(a,&b))|
  & Unwraps args,
    calls \lstinline|f|, wraps result in most restrictive label among all argument labels \\
Method call
  & \lstinline|mcall!(obj.m(a))|
  & Preserves receiver label $L$ on return value \\
Field access
  & \lstinline|mcall!(obj.field)|
  & Field access preserving label $L$ \\
\midrule
\multicolumn{3}{@{}l}{\textit{Implicit flow control}} \\
\addlinespace[2pt]
PC block
  & \lstinline|pc_block!{(L){...}}|
  & Initializes PC $= L$; rewrites all assignments to
    \lstinline|secure_assign_with_pc| which checks
    both explicit and implicit flows;
    update PC at nested branches \\
\midrule
\multicolumn{3}{@{}l}{\textit{Side-effect control}} \\
\addlinespace[2pt]
Attribute (fn)
  & \lstinline|#[side_effect_free_attr]|
  & Required attribute for functions in \lstinline|pc_block!| \\
Unchecked
  & \lstinline|unchecked_operation(e)|
  & Bypasses side-effect checks (escape hatch) \\
\bottomrule
\end{tabular}

\caption{\toolname APIs.}
\label{fig:programming-model}
\vspace{2em}
\end{figure}



\subsection{The Programming Model}
\label{sec:program_model}

\subsubsection*{Security Labels and Lattice}
\label{sec:labels}
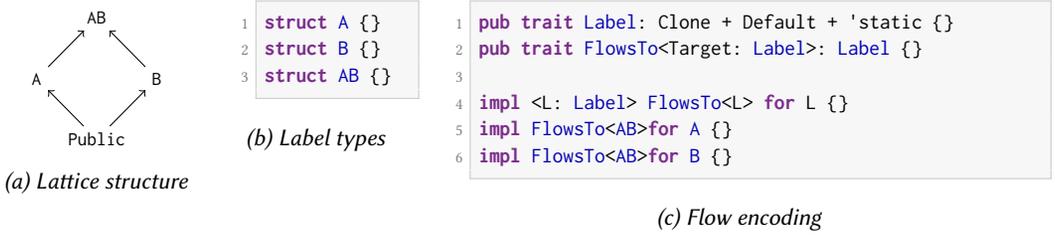
\begin{figure}[t]
    \centering
    \small 
    
    \begin{subfigure}[t]{0.19\columnwidth}
        \centering
        \vspace{0pt}
        \begin{tikzpicture}[scale=0.8, every node/.style={transform shape}]
            \node (AB)     at (1, 2)  {\texttt{AB}};
            \node (A)      at (0, 1)  {\texttt{A}};
            \node (B)      at (2, 1)  {\texttt{B}};
            \node (Public) at (1, 0)  {\texttt{Public}};
            \draw[->] (Public) -- (A);
            \draw[->] (Public) -- (B);
            \draw[->] (A)      -- (AB);
            \draw[->] (B)      -- (AB);
        \end{tikzpicture}
        \caption{Lattice structure}
    \end{subfigure}
    \hfill
    \begin{subfigure}[t]{0.18\columnwidth}
        \vspace{0pt}
\begin{lstlisting}[]
struct A {}
struct B {}
struct AB {}
\end{lstlisting}
        \caption{Label types}
    \end{subfigure}
    \hfill
    \begin{subfigure}[t]{0.58\columnwidth}
        \vspace{0pt}
\begin{lstlisting}[]
pub trait Label: Clone + Default + 'static {}
pub trait FlowsTo<Target: Label>: Label {}

impl <L: Label> FlowsTo<L> for L {}
impl FlowsTo<AB>for A {}
impl FlowsTo<AB>for B {}
\end{lstlisting}
        \caption{Flow encoding}
        \label{fig:lattice_relation}
    \end{subfigure}

    \caption{A four-element security lattice. (a) illustrates the lattice hierarchy ($\ell_1 \sqsubseteq \ell_2$). (b) shows the zero-sized types representing security levels. (c) demonstrates the encoding of the \lstinline|FlowsTo| relation using Rust's trait system.}
    \label{fig:flowsto}
\end{figure}

In \toolname, user specifies a pre-defined lattice in two aspects. First, security labels are defined as distinguished types. Moreover, the ordering on security labels is governed by the \lstinline|FlowsTo| trait, which is implemented for a label $\ell_1$ if and only if $\ell_1 \sqsubseteq \ell_2$ in the security lattice. For example, the Rust implementation in Figure~\ref{fig:flowsto} defines the ``diamond'' security lattice shown to the left. Note that since \toolname treats all ``unlabeled'' types as public data (i.e., label \cod{Public} in a lattice), we only need to define three non-public labels and their relation in Figure~\ref{fig:lattice_relation}.

Moreover, the join operation is defined as a trait that computes the least upper bound of two security labels at compile time, producing an associated output type \lstinline|Out| that represents the resulting security level. For identical labels, the join is idempotent (e.g., $A \sqcup A = A$), whereas for distinct labels, the result corresponds to their least upper bound (e.g., $A \sqcup B = AB$). A partial  implementation is shown below.
\begin{lstlisting}
pub trait Join<Other: Label>: Label {type Out: Label;}
impl Join<A> for A {type Out = A;}
impl Join<B> for A {type Out = AB;}
impl Join<AB> for A {type Out = AB;}
\end{lstlisting}

\subsubsection*{Labeled Data}
\label{sec:labeleddata}


In \toolname, all non-public data is wrapped in an abstract type \lstinline|Labeled<T, L: Label>|, where \lstinline|T| is the value type and \lstinline|L| is a security label as defined above. 

\begin{lstlisting}
pub struct Labeled<T, L: Label> {
    pub(crate) value: T,
    pub(crate) _marker: PhantomData<L>,
}
\end{lstlisting}
Note that \toolname employs a phantom-typed field, \lstinline|_marker|, to associate security labels with data. This design ensures that labeled values incur no runtime overhead. Furthermore, \toolname leverages the \lstinline|pub(crate)| visibility modifier to enforce opacity of these values, thereby restricting direct access. As a result, labeled values can only be manipulated through the interfaces provided by the \toolname library, which are described in subsequent sections.

\subsection{Explicit Flows}
\label{sec:explicit}

As introduced in Section~\ref{sec:denning}, an explicit flow occurs when an expression $e$ is assigned to a variable $x$. To enforce IFC, we must ensure that the security label of $x$ is at least as restrictive as the label of $e$. In a mainstream language like Rust, implementing this enforcement presents two primary challenges: type-system rigidity and annotation burden.

\subsubsection*{The Challenge: Type-System Rigidity}

The most significant hurdle in implementing Denning-style IFC within a standard type system is that security-safe flows often appear as type mismatches to the compiler. Consider the following example:

\begin{lstlisting}[language=Rust]
let mut x: i32 = 0; // Public
let mut y: Labeled<i32, A> = Labeled::new(10); // Secret
x = y; // (1) Rejected: Type mismatch
y = x; // (2) Rejected: Type mismatch
\end{lstlisting}

In assignment (1), Rust correctly prevents a security leak, but it does so because of a structural type mismatch between \lstinline|Labeled<i32, A>| and the primitive \lstinline|i32|, rather than by detecting an illegal information flow. However, in assignment (2), the same rigidity blocks a perfectly safe flow: moving public data to a secret variable ($\ell_{Public} \sqsubseteq \ell_{A}$).

This highlights a core limitation: a naive type-based approach is often too conservative to distinguish between a dangerous flow (High-to-Low) and a secure one (Low-to-High). To resolve this without intrusive compiler modifications, \toolname provides a unified approach using explicit relabeling and automatic inference.

\subsubsection*{Relabeling in Variable Assignment}

While a flow from label $\ell$ to $\ell'$ is secure whenever $\ell \sqsubseteq \ell'$, Rust rejects the assignment if $\ell \neq \ell'$. To bridge this gap, we introduce the \lstinline|relabel!(expr, L_target)| macro. This macro allows a programmer to explicitly ``upgrade'' the security label of an expression to satisfy the type system.

To ensure soundness, the macro invokes an internal function, \lstinline|__relabel_checked::<_, _, L_target>(...)|, which statically verifies the flow condition $\ell_{src} \sqsubseteq \ell_{target}$ via the \lstinline|L_src: FlowsTo<L_target>| trait (Section~\ref{sec:labels}). By using \lstinline|relabel!|, the programmer reduces the problem of arbitrary lattice flows to a standard, type-safe assignment:

\begin{lstlisting}
// Using the relabel! macro to satisfy the type system for line 4 above
y = relabel!(x, A); // Success: Static check confirms Public flows to A
\end{lstlisting}

This design ensures that label upgrades are explicit, sound, and checked entirely at compile-time. Furthermore, it allows \toolname to remain highly permissive while fully leveraging Rust's existing safety guarantees.

\subsubsection*{Label Propagation via Inference}

A key design goal of \toolname is to reduce the programmer's annotation burden, particularly for intermediate expressions such as \lstinline|x + y|. While \lstinline|relabel!| is necessary for cross-label assignments (partially due to assignment operator overloading is not allowed in Rust), \toolname automates label propagation during expression evaluation through operator overloading.

When an operation combines two values with labels $L_1$ and $L_2$, \toolname automatically computes the resulting label as the lattice join: \lstinline|Labeled<T, Join<L1, L2>>|. Because these resulting types are concrete and consistent with the security lattice, \toolname leverages Rust's native type inference engine to determine the labels of intermediate variables. 

\begin{lstlisting}[language=Rust]
let x: Labeled<i32, A> = Labeled::new(10);
let y: Labeled<i32, B> = Labeled::new(20);
let z = x + y; // z is automatically inferred as Labeled<i32, AB>
\end{lstlisting}

By utilizing the host language's inference engine, \toolname avoids the need for a specialized external constraint solver, such as the one employed by Jif~\cite{jif-cornell}. This allows for a more ergonomic programming model where explicit information flows are either automatically inferred or easily verified via relabeling.

\subsection{Implicit Flows}
\label{sec:implicit}
\subsubsection*{Challenges}
Explicit information flows between data are relatively straightforward to enforce, as type systems inherently check type compatibility between values. However, standard type systems typically do not track the control-flow context, represented by the $\pc$ label, which is required to enforce implicit flow checks (Section~\ref{sec:denning}). Consequently, the primary challenge is to track the $\pc$ label and incorporate it into type checking without modifying Rust’s underlying type system.

One key observation is that when the condition of a branch or loop is public, no additional mechanisms are required. The reason is that public values are represented using ordinary data types without security labels, and all implicit flows within the corresponding control structure are trivially secure because $\pc = \bot$ in this case. For example, consider 
\begin{lstlisting}
let x: bool = ...;
let y: Labeled<int32, A> = Labeled::new(10);
if x {
    y = relabel!(10, A);  // both implict flow and explit flow are secure
}
\end{lstlisting}
It is secure for the the type system to check only explicit flows at line 4 (as described in Section~\ref{sec:explicit}). This follows because (1) the type system requires the branch condition to be without a label, i.e., public, and (2) the implicit-flow constraint at line 4, $\pc \sqsubseteq \cod{A}$, trivially holds whenever $\pc=\bot$.

However, the Rust type system rejects any program whose branch or loop condition is non-public, i.e., has a labeled type, because labeled types are incompatible with the expected type \lstinline|bool|. On the positive side, this behavior preserves soundness without requiring additional mechanisms. The remaining challenge, however, is to make the type system more permissive in cases where the branch condition is non-public. To address the challenge, we introduce a novel macro \lstinline|pc_block!| to both track $\pc$ label and check implicit flows within the context.

\subsubsection*{PC Blocks}

For any branch or loop whose condition is non-public, \toolname requires an explicit annotation using the \lstinline|pc_block!| macro. At a high level, \lstinline|pc_block! {(L) {code}}| defines a lexical context in which the program counter label is initialized to $\ell$. Within this context, the macro enforces several restrictions: (1) it updates the $\pc$ label for each branch or loop encountered, (2) it performs implicit-flow checks for every assignment within the block, and (3) it ensures that any function or method call inside the block is free of side effects. We elaborate on each of these restrictions next.

\subsubsection*{Tracking PC Label}
The macro \lstinline|pc_block! {(L) {code}}| first initializes a local label, \lstinline|__pc|, to the annotated label \cod{L}. When the macro encounters a branch or loop statement, it rewrites the condition using \lstinline|inspect_condition()|, which extracts both the underlying boolean value and the security label of the condition. The program counter label is then \emph{joined} with the condition's label before entering the branch body via \lstinline|join_labels(__pc, _cond_label)|, which updates the $\pc$ for the encountered branch or loop at the type level. The resulting label \emph{shadows} the outer \lstinline|__pc| for the entire scope of the branch.

For example, the following code snippet illustrates the computed type-level \lstinline|__pc| associated with each branch, assuming that \lstinline|a| and \lstinline|b| have labels \lstinline|A| and \lstinline|B| respectively:
\begin{lstlisting}
pc_block! { (A) {    // label A
    if a {           // label A
        if b {       // label AB  
            ... 
        }            // pc reverts to A
}}}
\end{lstlisting}

\subsubsection*{Checking Implicit Flows}
Every assignment inside \lstinline|pc_block!| is rewritten to
\lstinline|secure_assign_with_pc|, which in addition to the explicit flow checks described in Section~\ref{sec:explicit}, also verifies that \lstinline|__pc: FlowsTo<L_target>|, ensuring that the target label is at least as restrictive than the current $\pc$.


%

For example, consider the following code snippet:
\begin{lstlisting}[language=Rust]
let secret: Labeled<bool, A> = ...;
let mut x: Labeled<int32, A> = Labeled::new(0);
let mut y: Labeled<int32, B> = Labeled::new(0);
pc_block! { (A)           // pc is set to A
  if secret {
    x = relabel!(1, A);   // type checking succeeds 
    y = relabel!(1, B);   // type checking fails
}}
\end{lstlisting}
The \lstinline|pc_block!| macro enforces implicit-flow checks by verifying the constraint \lstinline|A: FlowsTo<A>| at line 6. Since this constraint holds, the assignment at that line is well-typed. In contrast, the check at line 7, \lstinline|A: FlowsTo<B>|, fails; consequently, the type system correctly rejects the assignment at line 7.


\subsubsection*{Controlling Side Effects}

To control implicit flows via function calls inside a PC block, \toolname verifies that such functions are free of side effects. To do so, we follow Cocoon's implementation of the \lstinline|#[side_effect_free_attr]| attribute, and check that only functions with this verified attribute are used inside a PC block. 
%
As the verification of the \lstinline|#[side_effect_free_attr]| annotation follows the same approach as in Cocoon, we refer interested readers to their paper~\cite{lamba2024} for further details.

Similar to Cocoon, \toolname provides an interface, \lstinline|unchecked_operation|, as an escape hatch that bypasses all security checks. This interface is intended to address potential imprecision in the verification of the \lstinline|#[side_effect_free_attr]| annotation, as well as to support intentional information release via side effects. Similar to declassification interfaces such as \lstinline|declassify(x)|, the programmer remains responsible for ensuring that the use of such unchecked operations does not introduce unintended information leakage. Notably, we observe that \lstinline|unchecked_operation| is not used in any of the case studies presented in our evaluation (Section~\ref{sec:evaluation}), suggesting that our approach is sufficiently permissive in practice.

\subsubsection*{\cod{pc\_block} vs. \cod{secret\_block}}

While PC blocks in \toolname shares certain mechanisms with secret blocks in Cocoon, there are several notable differences. The primary distinction is that \cod{pc\_block} annotations are required only for branches or loops with non-public conditions, whereas \cod{secret\_block} must be applied to \emph{any code} that manipulates secret values. As a result, we observe a substantial reduction in annotation overhead: \toolname requires only 3 PC blocks, compared to 36 secret blocks in Cocoon (Section~\ref{sec:annotation}). As a consequence, fewer functions requires the \lstinline|#[side_effect_free_attr]| annotations overall. Furthermore, aside from the verification of side-effect-free functions, the additional mechanisms (including tracking PC label and checking implicit flows) enforced by \lstinline|pc_block!| are not present in Cocoon.

\subsection{Third-Party Security-Agnostic Functions and Methods}
\label{sec:funccall}

\subsubsection*{Challenges}
As discussed in Section~\ref{sec:func_calls_mot}, nontrivial programs routinely depend on the standard Rust library, built-in methods (e.g., \lstinline|.len()|, \lstinline|.contains()|, and \lstinline|.push()|), and third-party libraries (e.g., \lstinline|str::parse|). However, these methods and functions are \emph{security-agnostic}, in the sense that they operate on raw types \lstinline|T| rather than labeled types \lstinline|Labeled<T, L>|. Hence, programs that manipulate labeled values (e.g., values of type \lstinline|Labeled<String, A>|) cannot directly reuse such functions without rewriting them to be label-aware, which is generally infeasible.

In the Cocoon programming model, third-party libraries are treated as trusted (i.e., part of the trusted computing base), alongside the Rust compiler and the standard Rust library. Cocoon's procedural macro maintains an ``allowlist'' of functions that are deemed side-effect free by its designers. To address the type incompatibility described above, Cocoon adopts the following pattern when invoking trusted libraries: sensitive arguments are first declassified to match the expected function signatures, and the return values are subsequently re-wrapped with appropriate security labels. However, as noted in Section~\ref{sec:func_calls_mot}, this pattern in prone to misuses, as declassified values may bypass all subsequent IFC enforcement mechanisms.

In the \toolname programming model, we adopt the same threat model: third-party libraries are treated as trusted, alongside the Rust compiler and the standard library. However, to avoid the error-prone patterns employed by Cocoon, \toolname treats each function or method in trusted libraries as a \emph{black box}. In particular, it automatically computes a security label that is at least as restrictive as the labels of all input parameters, and assigns this label to the function’s return value. This behavior is enforced through the macros \lstinline|fcall!| and \lstinline|mcall!|, which we describe next.

\subsubsection*{Invocation Macros in \toolname.}
\toolname introduces the \lstinline|fcall!| macro to enable the use of trusted, security-agnostic Rust functions with labeled data. Concretely, the macro operates as follows. First, it generates a sequence of nested \lstinline|__chain| or \lstinline|__chain_ref| calls, one per argument. Each closure unwraps a \lstinline|Labeled| value and binds the underlying data to a fresh variable, thereby exposing only raw values within the innermost scope. Second, at the innermost level, the original function \lstinline|f| is invoked on the unwrapped arguments. Third, the result is initially wrapped as \lstinline|Labeled<R, Public>|; subsequently, the enclosing \lstinline|__chain| calls propagate and join the labels of all arguments via the \lstinline|Join| trait. As a result, the final output is associated with the \emph{most restrictive label} among all input parameters, ensuring that the resulting label conservatively reflects the sensitivity of the inputs.

The definition of \lstinline|fcall!(f(a, b))| is shown below:
\begin{lstlisting}
{
    use typing_rules::function_rewrite::{SecureChain};
    (a).chain(|__v0| {
        (b).chain(|__v1| {
            Labeled::<_, Public>::new(f(__v0, __v1))
        })
    })
}
\end{lstlisting}




For method calls (e.g., \lstinline|obj.method(args)|) and field accesses (e.g., \lstinline|obj.field|), both of which require accessing the receiver’s underlying value, \toolname introduces the \lstinline|mcall!| macro to ensure secure usage. The design of \lstinline|mcall!| follows the same general principles as \lstinline|fcall!|; however, instead of computing the most restrictive label among all arguments, \lstinline|mcall!| \emph{preserves} the label of the receiver. This design effectively treats each object as a black box for the purposes of information-flow control. Consequently, the result inherits the receiver's label, ensuring that sensitivity is determined solely by the object being accessed. This is achieved with a label-preserving helper function below:
\begin{lstlisting}
fn __mcall_preserve_label<T, U, L: Label>(wrapper: &Labeled<T, L>,func: impl FnOnce(&T) -> U) -> Labeled<U, L> {
    Labeled::<U, L>::new(func(wrapper.inner()))
}
\end{lstlisting}
This helper function borrows the underlying value, applies the method or field access within a closure, and wraps the result in a new \lstinline|Labeled| value that preserves the original label~\lstinline|L|. For example:
\begin{lstlisting}
mcall!(placement.start_row) // Field access
// expands to:
__mcall_preserve_label(&placement, |inner| inner.start_row)
\end{lstlisting}
Furthermore, a single \lstinline|mcall!| invocation can express a chain of method calls on a labeled receiver. The macro recursively deconstructs the chain of \lstinline|mcall!| expressions to identify the root labeled receiver, and then reconstructs the entire chain as the body of a single closure passed to \lstinline|__mcall_preserve_label|. For example:
\begin{lstlisting}
mcall!(key.chars().all(f))
// expands to:
__mcall_preserve_label(&key, |inner| inner.chars().all(f))
\end{lstlisting}
The entire call chain is evaluated within the closure on the unwrapped inner value, and only the final result is re-wrapped. As a consequence, intermediate values (e.g., the \lstinline|Chars| iterator produced by \lstinline|.chars()|) are never individually labeled. This design is sound because the closure boundary ensures that such intermediate values remain unobservable outside the macro, while the final result preserves the label of the original receiver.

\subsection{Limitation of \toolname}
While \toolname provides a robust framework for static, lattice-based information flow control, several security dimensions remain outside the current scope of this work. Because \toolname enforces IFC statically, security policies are resolved entirely at compile time; consequently, we do not support dynamic labels where the secrecy level of a value changes at runtime~\cite{li2022}. Furthermore, this work focuses strictly on explicit and implicit flows, so side-channel protections (such as timing channels) are not currently addressed. While \toolname's design follows established type-based IFC principles, a formal proof of noninterference for this specific Rust implementation is beyond the scope of this paper.

The current design of \toolname utilizes a predefined lattice, which simplifies the implementation of the \lstinline|FlowsTo| and \lstinline|Join| traits but limits the system to policies where the security lattice is static. Additionally, the program counter ($pc$) label for \lstinline|pc_block!| must currently be specified manually by the programmer at each call site. We intend to explore automatic $pc$ inference in future iterations to reduce this manual overhead. Moreover, because our \lstinline|pc_block!| implementation inherits properties from Cocoon's \lstinline|secret_block!|, only functions within a predefined allowlist can be safely invoked within these blocks to prevent side effects. A more scalable solution would involve a custom procedural macro to statically verify the side-effect-free nature of arbitrary Rust functions.

Finally, \toolname's current model assumes a single-threaded execution context. Extending \toolname to handle concurrent Rust, where data-race freedom must be reconciled with information flow across multiple threads, remains an open research challenge.

\section{Evaluation}
\label{sec:evaluation}

We implement \toolname as a library on top of Rust 1.69 to make it directly comparable with Cocoon, which is implemented on the same Rust version. Our implementation consists of about 2000 lines of Rust code. All experiments were conducted on a dual-socket Intel Xeon 6517P server, 251 GB RAM, running Ubuntu 24.04.3 on Rust version 1.69. We open-sourced \toolname at~\cite{implementation}\footnote{Anonymized for peer review.}.

To evaluate the usability and performance of \toolname, we re-implemented all four applications from the Cocoon code repository, including Calendar, Battleship, Spotify TUI, and Servo, as well as an additional application, JPMail~\cite{hicks2006}, a substantial information flow case study originally implemented in Jif. For each case study, we translate the original Rust code from the Cocoon repository, or the Jif implementation in the case of JPMail, into its \toolname equivalent, while enforcing the same IFC policies as in their respective Cocoon and Jif implementations. 


\begin{itemize}
    \item Calendar is a simple program that determines the mutual availability of Alice and Bob, where each person’s calendar is considered confidential.
    
    \item Battleship is a classic two player guessing game in which each player privately places a fleet of ships on a hidden grid. Players take turns guessing coordinates on their opponent’s grid, and the first player to sink all opposing ships wins. This example has been used as a case study in Jif and its extension JRIF~\cite{jif-cornell, kozyri2020}. 

    \item Spotify TUI is an open-source terminal-based Spotify client written in Rust, allowing users to browse playlists, search for tracks, control playback, and manage their Spotify library entirely from the command line. This application processes user credentials, including user IDs and passwords. So any information derived from or related to the password is treated as confidential.

    \item Servo is Mozilla's browser engine implemented in Rust. Following the Cocoon implementation, we enforce the same security policy under which JavaScript programs may only access HTTP responses originating from the same server. Responses from different origins are treated as opaque and must not be used exclusively.

    \item JPMail is a secure email system that uses IFC to protect email confidentiality at the language level. It was originally written in Jif, enabling users to send, receive and reply to emails while preserves privacy. The password, content of the email, and Private keys are considered secrets in this application. 
\end{itemize}

Across the five case studies, we observe that although Spotify TUI and Servo comprise the largest codebases (11K and 327K LoC, respectively), their security critical components are relatively small. In contrast, smaller applications such as JPMail (2K LoC) require more extensive security annotations due to more intricate security-critical code throughout. Nevertheless, Spotify TUI and Servo illustrate realistic settings in which third party libraries constitute the majority of the codebase.

In this section, we address the following research questions. Whenever possible, we also compute the same metrics for the Cocoon equivalents to enable a fair comparison.
\begin{description}
    \item[RQ1 Code Annotation Burden:] How much annotation burden is required to enforce IFC with \toolname? More specifically, how many security related type annotations and macros are needed to verify IFC policies in \toolname?
    
    \item[RQ2 Permissiveness:] How often must escape hatches, including \lstinline|declassify| (excluding intentional uses, such as releasing the mutually available dates of Alice and Bob in the Calendar case study) and \lstinline|unchecked_operation|, be used in \toolname to bypass security checks when the security enforcement mechanism proves too restrictive in practice?
    
    \item[RQ3 Compilation Overhead:] What are the compilation time overhead of the IFC enforcement mechanisms? Note that \toolname enforces IFC entirely at compile time; consequently, verified Rust programs incur no runtime overhead by design.
\end{description}

By re-implementing these examples, we demonstrate that \toolname achieves the same security objectives as Cocoon using a principled Denning-style approach while requiring fewer modifications to the original source code. The primary distinction is that Cocoon relies on \lstinline|secret_block!| constructs, whereas \toolname associates labels directly with individual variables. While variable-level labeling is often perceived to increase the annotation burden, we show that our use of label inference largely mitigates this overhead. Furthermore, the permissive programming model of \toolname significantly reduces the number of unnecessary declassification and unchecked operations that can potentially expose secret values to the public.

\subsection{Code Annotation Burden}
\label{sec:annotation}

\begin{table}[]
\centering
\caption{System Comparison: LoC and Annotation Density per Tool}
\label{tab:ifc-loc}
\small
\begin{tabular}{l r rrr rrr} 
\toprule
& \textbf{Original} & \multicolumn{3}{c}{\textbf{Filament}} & \multicolumn{3}{c}{\textbf{Cocoon}} \\
\cmidrule(lr){2-2} \cmidrule(lr){3-5} \cmidrule(lr){6-8}
\textbf{Project} & \textbf{LoC} & \textbf{LoC} & \textbf{Label} & \textbf{API} & \textbf{LoC} & \textbf{Label} & \textbf{API} \\
\midrule
Calendar     & 29      & 36 (+24.1\%)     & 14 & 2 & 48 (+65.5\%)     & 0  & 26 \\
Battleship   & 286     & 293 (+2.4\%)    & 16 & 2 & 315 (+10.1\%)    & 11  & 23\\
Spotify TUI  & 10,912  & 10,966 (+0.5\%)  & 36 & 23 & 11,029 (+1.1\%) & 13 & 49 \\
Servo        & 327,096 & 327,160 (+0.02\%) & 61  & 1 & 327,129 (+0.01\%) & 64 & 1  \\
JPMail       & 1,894   & 2,006 (+5.9\%)  & 199 & 53 & ---     & ---& --- \\
\bottomrule
\end{tabular}
\end{table}

A major concern for any language-based IFC tool is the extent of code modification required for verification, including both security-related type annotations and IFC-library API calls. To estimate the annotation burden associated with \toolname and Cocoon, we summarize the total lines of code (LoC) and annotations (categorized into label annotations and API calls) in Table~\ref{tab:ifc-loc}.

For the first four applications, where the original Rust implementations (without IFC) are available, both \toolname and Cocoon exhibit similar LoC, with only minor to modest increases---except for Calendar, due to its small codebase. The additional lines primarily occur at framework boundaries that the label system cannot transparently cross. For example, because the \lstinline|Serialize| and \lstinline|Deserialize| traits used in Servo are not aware of security types such as \lstinline|Labeled<T, L>|, auxiliary helper functions are required in Spotify TUI for both \toolname and Cocoon. Nevertheless, the majority of application logic, including UI rendering, protocol parsing, and game loops, remains unchanged. This suggests that, in both systems, annotation effort is localized to security-critical regions. This also explains why the LoC increase is not strictly proportional to the size of the original codebase. For JPMail, originally implemented in Jif with built-in IFC support, we observe a similarly modest increase, likely due to Jif's use of a customized type system.

Compared to Cocoon, \toolname requires fewer lines of code for Battleship (293 vs.\ 315), Calendar (36 vs.\ 48), and Spotify TUI (10,966 vs.\ 11,029), while it is slightly larger for Servo (327,160 vs.\ 327,129). Overall, \toolname implementations tend to be more concise, particularly in the percentage increase over the original code.

To better understand the extent of changes introduced by security enforcement, we count the total number of lines containing at least one security-related annotation. These include label annotations at variable declarations and function signatures, as well as uses of \toolname interfaces (Figure~\ref{fig:programming-model}). We apply the same methodology to Cocoon and summarize the results in the ``Label'' and ``API'' columns of Table~\ref{tab:ifc-loc}.

Because \toolname adopts a fine-grained, Denning-style approach, a higher number of label annotations is expected. As shown in Table~\ref{tab:ifc-loc}, the annotation counts are higher than Cocoon's for all programs except Servo. Nevertheless, the number of lines requiring security label annotations in \toolname remains low in practice, with only 16 lines for Battleship, 36 for Spotify TUI, and 61 for Servo. Only JPMail incurs a substantially higher burden, as many data sources must be labeled as secret and a large portion of its codebase is security-critical. Overall, the label annotation burden in \toolname is moderate at most; this is notable given that \toolname enforces both explicit and implicit flows at the expression level while fully leveraging Rust's inference engine.

A clear strength of \toolname compared to Cocoon is the significant reduction in IFC library API calls. The permissive programming model of \toolname completely eliminates the need for Cocoon's \lstinline|secret_block!|. Consequently, this reduces: (1) the necessity for frequent declassification calls and attribute tags; and (2) the repetitive overhead of \lstinline|wrap| and \lstinline|unwrap| calls within every \lstinline|secret_block!|, as discussed in Section~\ref{sec:uncheckedfunction}. The API calls in \toolname consist mainly of macros like \lstinline|fcall!| and \lstinline|mcall!| to interact with third-party libraries, along with several \lstinline|#[side_effect_free_attr]| attributes for functions executed under PC blocks.

A breakdown for each application is shown in Table~\ref{tab:ifc-annotations}. Across all four applications, the annotation burden for \toolname remains low: Battleship uses only 2 API calls, while Spotify TUI requires 23 (7 \lstinline|declassify|, 14 \lstinline|fcall!|/\lstinline|mcall!|, 1 \lstinline|relabel!|, and 1 \lstinline|pc_block!|). Calendar and Servo require the least instrumentation. JPMail is the most annotation-heavy, with 12 declassification calls reflecting its complex multi-principal security policy; these calls occur mostly at the socket interface. The \lstinline|fcall!| and \lstinline|mcall!| macros propagate labels through function calls without requiring manual argument declassification, reducing the total annotation count compared to a block-based approach.

\begin{table}
\centering
\caption{API usage in \toolname}
\label{tab:ifc-annotations}
\small
\begin{tabular}{l r r r r r r }
\toprule
\textbf{Project} & \texttt{declassify} & \texttt{fcall!/mcall!} & \texttt{pc\_block!} & \texttt{relabel!} \\
\midrule
Calendar        & 1 & 0  & 1 & 0  \\
Battleship & 2 & 0  & 0 & 0  \\
Spotify TUI     & 7 & 14  & 1 & 1 \\
Servo           & 1 & 0  & 0 & 0 \\
JPMail          & 12 & 13  & 1 & 0 \\
\bottomrule
\end{tabular}
\end{table}

Another factor contributing to the reduced API call count is \toolname's more permissive programming model, which avoids the need for spurious escape hatches (such as \lstinline|unchecked_operation| and \lstinline|declassify|) required by Cocoon to bypass security checks. We elaborate on this point in the following discussion.


\subsection{Permissiveness}
As discussed in Section~\ref{sec:uncheckedfunction} and Section~\ref{sec:func_calls_mot}, a limitation of the Cocoon programming model is that programmers are sometimes required to use escape hatches, such as \lstinline|unchecked_operation| and \lstinline|declassify|, either to circumvent checks that are overly restrictive in practice or to accommodate coding patterns necessary for reusing third-party libraries. To better understand the frequency and underlying reasons for these escape hatches, we measure the number of invocations for these APIs. Additionally, we count the number of \lstinline|pc_block!| and \lstinline|secret_block!| instances in \toolname and Cocoon, respectively. The results are summarized in Table~\ref{tab:api-usage}.

\begin{table}[!htb]
\centering
\begin{threeparttable}
\caption{API usage comparison: Filament vs. Cocoon.}
\label{tab:api-usage}
\setlength{\tabcolsep}{15pt} 
\small
\begin{tabular}{l ccc ccc}
\toprule
& \multicolumn{3}{c}{\textbf{Filament}} & \multicolumn{3}{c}{\textbf{Cocoon}} \\
\cmidrule(lr){2-4} \cmidrule(lr){5-7}
\textbf{Project} & {\textbf{D}} & {\textbf{U}} & {\textbf{P}} & {\textbf{D}} & {\textbf{U}} & {\textbf{S}} \\
\midrule
Calendar     & 1 & 0 & 1 & 1  & 0 & 17 \\
Battleship   & 2 & 0 & 0 & 2  & 2 & 4  \\
Spotify TUI  & 7 & 0 & 1 & 17 & 0 & 15 \\
Servo        & 1 & 0 & 0 & 1  & 0 & 0  \\
\bottomrule
\end{tabular}
\begin{tablenotes}
  \item APIs: \textbf{D}--\lstinline|declassify|, \textbf{U}--\lstinline|unchecked_operation|, \textbf{P}--\lstinline|pc_block!|, \textbf{S}--\lstinline|secret_block!|
\end{tablenotes}
\end{threeparttable}
\end{table}

The \lstinline|unchecked_operation| API, which completely bypasses all security checks in both Cocoon and \toolname, is used twice in Cocoon's implementation of Battleship, whereas the \toolname implementation does not require it. This difference arises because \toolname's Denning-style analysis eliminates unnecessary secret blocks, thereby avoiding superfluous side-effect checks on certain functions, as illustrated in Figure~\ref{fig:check_guess_f}.
Figure~\ref{fig:unchecked_rand_c} shows another case where Cocoon's \lstinline|random| function requires \lstinline|unchecked_operation| because it is used within another function tagged with \lstinline|#[side_effect_free_attr]| (\lstinline|random_maybe_illegal_placement|), which is used in a secret block. \toolname avoids this issue because \toolname can distinguish implicit flows from explicit ones, and no pc block is required in the Battleship application. Hence, no side effect checks are required.

Although only two instances of \lstinline|unchecked_operation| appear across the four applications in the Cocoon repository, this scarcity is likely due to either the small scale of two applications (Battleship and Calendar) or a reliance on trusted third-party libraries (Spotify TUI and Servo), which eliminates the need for side-effect checks under the assumed threat model. For larger applications such as JPMail, we expect patterns similar to those in Battleship to occur more frequently, leading to a greater reliance on \lstinline|unchecked_operation|. In contrast, \toolname mitigates this limitation by explicitly checking information flows between secret data and target sinks, requiring \lstinline|unchecked_operation| only when a function with side effects is invoked within a sensitive branch. 

\begin{figure}[t]
    \centering
    \begin{subfigure}{\textwidth}
\begin{lstlisting}[]
#[side_effect_free_attr]
pub fn random(limit: usize) -> usize {
    (*@\highlightlineCo{unchecked\_operation}@*)unchecked_operation(rand::thread_rng().gen_range(0..limit))
}

#[side_effect_free_attr]
fn random_maybe_illegal_placement(grid: &Grid<bool>, ship: &Ship) -> Placement {
    let orientation = util::random(2);
    if orientation == 1usize { /* ... */ } else { /* ... */ }
    let row = util::random(row_limit);
    let col = util::random(col_limit);
    Placement { /* ... */ }
}
\end{lstlisting}
    \caption{\lstinline|random| function in Cocoon, utilizing \lstinline|unchecked_operation|.}
    \label{fig:unchecked_rand_c}
    \end{subfigure}

    \vspace{1em} 

    \begin{subfigure}{\textwidth}
\begin{lstlisting}[]
pub fn random(limit: usize) -> usize {
    rand::thread_rng().gen_range(0..limit)
}

fn random_maybe_illegal_placement(grid: &Grid<bool>, ship: &Ship) -> Placement {
    /* ... same logic ... */
}
\end{lstlisting}
    \caption{\lstinline|random| function in \toolname, without \lstinline|unchecked_operation|.}
    \label{fig:unchecked_rand_f}
    \end{subfigure} 

    \caption{Comparison of the \lstinline|random| function implementation in the Battleship case study.}
    \label{fig:unchecked_rand}
\end{figure}

The second escape hatch is \lstinline|declassify|, available in both tools. Since intentional declassification is required in many applications (e.g., releasing mutually available dates in Calendar), we manually examined all 11 occurrences of \lstinline|declassify| in the \toolname implementations. We find that all such uses are intentional and necessary for application functionality. In contrast, Cocoon's 10 additional uses (all within Spotify TUI) are unnecessary, potentially exposing secret values to the public. We have already discussed a representative example of this category in Figure~\ref{fig:set-device-id}.

We attribute the increased permissiveness of the \toolname programming model primarily to the significant reduction in secret blocks compared to Cocoon. In total, 36 secret blocks are used across the Cocoon applications. Because Cocoon does not distinguish between explicit and implicit flows, IFC enforcement within these blocks must be conservative, disallowing side effects to prevent implicit flows. In contrast, \toolname eliminates the need for secret blocks by design, applying similar restrictions only within \lstinline|pc_block!|. Notably, this construct is required only for branches conditioned on sensitive values and is used only once in Spotify TUI. A second contributing factor is the introduction of the \lstinline|mcall!|, \lstinline|fcall!|, and \lstinline|relabel!| macros, which eliminate the need to declassify labeled parameters when invoking third-party library functions.

Moreover, because Rust supports label inference, \toolname does not require wrapping all secret computations within a single block, allowing the code to follow the original structure faithfully. Across our case studies, we wrap sensitive values in \lstinline|Labeled<T, L>| upon declaration and propagate them using our macros to enforce label-safe operations. Explicit \lstinline|declassify| calls are inserted only where the logic genuinely requires revealing secret data, and \lstinline|pc_block!| is utilized only when control flow branches on secret values.



\subsection{Compilation Overhead}


Since \toolname relies on procedural macros, it generates auxiliary code that must be processed by the compiler. In particular, \lstinline|pc_block!| produces a dual path expansion similar to Cocoon to check if there is information leak through implicit flow statically, while \lstinline|fcall!| and \lstinline|mcall!| introduce additional scaffolding to mediate library calls. Although these constructs are eliminated during optimization, they still need to be parsed and type checked. Accordingly, we evaluate whether such expansions have a significant impact on compilation time and compare the performance of \toolname with that of Cocoon.

\begin{table}[!htb]
\centering
\caption{Compilation time comparison: \toolname vs. \textsc{Cocoon} (relative to baseline)}
\label{tab:performance-delta}
\small
\begin{tabular}{l r rr rr}
\toprule
& \multicolumn{1}{c}{\textbf{Baseline}} & \multicolumn{2}{c}{\textbf{\toolname}} & \multicolumn{2}{c}{\textbf{\textsc{Cocoon}}} \\
\cmidrule(r){2-2} \cmidrule(lr){3-4} \cmidrule(lr){5-6}
\textbf{Project} & \textbf{Median (s)} & \textbf{Median (s)} & \textbf{Overhead} & \textbf{Median (s)} & \textbf{Overhead} \\
\midrule
Spotify TUI & 22.20  & 22.30  & +0.45\%  & 22.43  & +1.04\%  \\
Servo       & 193.86 & 189.36 & --2.32\% & 195.56 & +0.88\%  \\
Battleship  & 4.69   & 5.15   & +9.81\%  & 5.71   & +21.75\% \\
Calendar    & 4.14   & 4.25   & +2.66\%  & 4.97   & +20.05\% \\
\bottomrule
\end{tabular}
\end{table}

To evaluate this overhead, we measured median build times for each project.
Table~\ref{tab:performance-delta} reports the results for clean builds. The data demonstrates that \toolname's instrumentation introduces negligible overhead for large, dependency-heavy projects. Although Battleship and Calendar exhibit a more noticeable relative overhead, we attribute this to their small baselines. For example: Battleship has few external dependencies, the fixed cost of compiling the IFC library crates (\cod{macros} and \cod{typing\_rules}) represents a larger share of the total build time. Conversely, for Spotify TUI and Servo, this fixed cost is amortized over the compilation of heavyweight dependencies, making the overhead nearly unmeasurable. Spotify TUI shows a negligible difference (0.45\%), while Servo compiles slightly faster under \toolname ($-$2.32\%), a variance we attribute to build-cache fluctuations rather than a genuine speedup.

These results, shown in Table~\ref{tab:performance-delta}, confirm that \toolname imposes minimal compilation overhead. Despite using a more granular Denning-style approach, the overhead remains lower than that of Cocoon in all cases. This suggests that the overhead scales with the ratio of IFC library compilation to the total build time rather than with the volume of annotated code. Consequently, as a project grows in complexity, the relative overhead of adopting \toolname diminishes further. 

\section{Related Work}


\subsubsection*{Granularity of IFC}
The distinction between fine-grained and coarse-grained systems represents a fundamental design choice in IFC. Fine-grained type systems, such as Jif~\cite{myers1999} and FlowCaml~\cite{flowcaml}, track security labels at the level of variables and expressions. In contrast, coarse-grained systems, commonly used in IFC operating systems~\cite{krohn2007information, zeldovich2008securing}, associate labels with larger entities such as subjects and objects (e.g., processes and files). Several Haskell-based IFC systems built on the LIO model~\cite{buiras2015hlio, heule2015ifc, stefan2011, vassena2018mac, polikarpova2020} adopt an intermediate approach by tracking labels at the granularity of code blocks, which we refer to as block-level IFC in this paper. Although Cocoon is implemented in Rust, it conceptually also provides block-level IFC (Section~\ref{sec:blocklevel}).

The choice of granularity has been widely studied in the IFC literature. Rajani et al.~\cite{rajani2017type} identify granularity as a key factor in the design of IFC type systems. In subsequent work~\cite{rajani2018types}, they show that static fine-grained and static coarse-grained systems are equally expressive despite their structural differences, providing both a formal semantics and a type-preserving translation between the two. Similar discussions have arisen in the context of dynamic IFC for systems and programming languages~\cite{roy2009laminar, stefan2011}. This line of work culminates in Vassena et al.~\cite{vassena2019fine}, which demonstrates that fine-grained and coarse-grained dynamic systems are also equally expressive.

Although fine-grained and coarse-grained IFC systems are theoretically equivalent in expressiveness, the choice of granularity in a programming model remains an important design consideration. As argued throughout this paper, developing IFC-enabled programs in \toolname is more intuitive and natural for programmers. Moreover, the resulting code remains closer to conventional Rust code written without IFC, which facilitates porting existing Rust projects to \toolname.

\subsubsection*{IFC Tools and Application}

Prior IFC tools span both language extensions and library-based approaches. Early systems such as Jif~\cite{jif-cornell} and Flow Caml~\cite{flowcaml} provide fine-grained IFC for Java-like and OCaml-like languages, respectively, by extending their compilers and type systems. While these designs offer strong security guarantees, they rely on nonstandard compilation toolchains, which has limited the practical adoption of these otherwise well-founded techniques.

To improve accessibility for programmers, subsequent research explored embedding IFC frameworks directly within existing programming languages. Many of these efforts target functional languages, as they have expressive built-in type systems. For example, Haskell-based systems like LIO (Labeled IO)~\cite{stefan2011} and HLIO~\cite{buiras2015hlio} enforce dynamic IFC through a security monad. In these designs, security labels are associated with monadic computations rather than individual values, an approach initially explored by Li and Zdancewic~\cite{li2006encoding} and later simplified in Russo’s lightweight Haskell IFC library~\cite{russo2008library}, where sensitivity is tracked at the boundaries of monadic blocks.
More recent work has extended this line of research by incorporating stronger static guarantees. For instance, Lifty~\cite{polikarpova2020} integrates the LIO model with liquid types to statically verify IFC policies and automatically repair policy violations. Similarly, DepSec~\cite{gregersen2019dependently} demonstrates that dependently typed languages such as Idris can enforce data-dependent security policies entirely at the type level.

Besides functional settings, researchers have also explored embedding IFC directly within existing \emph{imperative programming languages}. One common approach is to extend existing compilers with an information-flow analysis pass that tracks dependencies and enforces security policies during compilation. Representative systems following this strategy include Flowistry~\cite{crichton2022modular}, Pidgin~\cite{pidgin}, FlowFence~\cite{fernandes2016flowfence}, and CtChecker~\cite{zhou2024}.
However, such systems require customized compilers, making them less compelling for end users. Moreover, maintaining customized compilers to keep pace with the evolution of their underlying base languages can be both costly and time-consuming. 

More recently, Cocoon~\cite{lamba2024} and its successor Carapace~\cite{beardsley2025carapace}, which \toolname is inspired by, present the first static IFC systems that work with a mainstream imperative language without modifying the compiler. As discussed in Section~\ref{sec:blocklevel}, both systems provide block-level IFC, whereas \toolname\ supports Denning-style, fine-grained IFC. Carapace additionally supports dynamic labels, a feature that is currently absent in \toolname. This represents an orthogonal direction that we plan to explore in future extensions of \toolname.

\section{Conclusion and Future Work}
We presented \toolname, a Denning-style static information-flow control (IFC) library for Rust that requires no compiler modifications. In contrast to block-level IFC approaches such as Cocoon, \toolname enforces both explicit and implicit flows at the granularity of expressions and variables. This design allows values at various security labels to coexist within the same scope while still ensuring end-to-end security through Denning-style IFC enforcement.

Through its novel design, \toolname addresses three key challenges in constructing a practical IFC library for Rust. First, it enforces explicit-flow checks at the granularity of expressions and variables, yielding a more permissive programming model without requiring substantial code rewriting compared to prior work, while also reducing manual annotations by leveraging Rust’s type inference system. Second, the novel \lstinline|pc_block!| construct enables implicit-flow enforcement, a defining feature of Denning-style IFC systems, that is absent in prior approaches without compiler modifications. Third, the \lstinline|fcall!| and \lstinline|mcall!| macros provide seamless interoperability with the standard library and third-party functions, without introducing error-prone patterns that are susceptible to misuse.

Our evaluation demonstrates that \toolname incurs negligible compilation overhead and requires only modest annotation effort, particularly in programs that heavily depend on external libraries. Furthermore, \toolname offers a more permissive programming model than Cocoon: implementations written using \toolname rely less frequently on escape hatches that bypass security checks compared to their Cocoon counterparts.

For future work, we plan to extend \toolname to support dynamic IFC policies, in which the sensitivity and integrity of data may evolve over time. We also aim to investigate the scalability of \toolname to larger codebases and to explore whether its macro-based design can be further streamlined through tighter integration with Rust’s procedural macro ecosystem.

\section*{Data Availability Statement}
An initial artifact for \toolname consists of the core IFC library and the complete source code for the examples discussed in this paper. The artifact is currently compatible with Rust 1.69, with support extending up to Rust 1.83. Please note that compilation times may vary depending on the hardware specifications of the host machine.

We intend to submit the full artifact for Artifact Evaluation. A temporary version of the source code, including the implementation of both \toolname and the examples is available at the following link~\cite{implementation}.

\bibliographystyle{ACM-Reference-Format}
\bibliography{ref}

\clearpage
\appendix

\end{document}